\documentclass[12pt]{article}
\title{Black hole information paradox without Hawking radiation}
\author{Hrvoje Nikoli\'c \\
Theoretical Physics Division, Rudjer Bo\v{s}kovi\'{c} Institute, \\
P.O.B. 180, HR-10002 Zagreb, Croatia \\
{\normalsize e-mail: hnikolic@irb.hr} \\
\makebox[1in]{} \\
}
\date{\today}
\begin{document}
\maketitle
\begin{abstract}
By entangling soft massless particles one can create an arbitrarily large amount of entanglement entropy 
that carries an arbitrarily small amount of energy. Dropping this entropy into the black hole (b.h.) one can
increase the b.h. entropy by an amount that violates Bekenstein bound or any other reasonable bound,
leading to a version of b.h. information paradox that does not involve Hawking radiation. 
Among many proposed solutions of the standard b.h. information paradox with Hawking radiation, only a few 
can also resolve this version without the Hawking radiation. The assumption that both versions
should be resolved in the same way significantly helps to reduce the space of possible resolutions.
\end{abstract}
\vspace*{0.5cm}


\section{Introduction}

The prediction of Hawking radiation \cite{hawk1} leads to the black hole (b.h.) 
information paradox \cite{hawk2}. 
Namely, the final state of Hawking radiation 
appears to be a mixed thermal state, and purity of the full state cannot be encoded 
into entanglement between radiation and b.h. degrees of freedom, either because 
the b.h. evaporated completely, or the radiation contains more entropy than can be 
stored into the b.h. with a limited entropy capacity. Thus it looks as if the full state
of the universe evolves from a pure initial state without Hawking radiation into a 
mixed state with radiation, which contradicts the general principle of unitarity, according to which
pure states of closed quantum systems can evolve only into pure states.
There are many proposals for possible resolutions of the paradox
as reviewed e.g.~in \cite{gid,har,pres,pag,gid2,str,math1,math2,hoss,dundar,harlow,polchinski,chakra,marolf,fabbri},
for example that the principle of unitarity is not valid in quantum gravity, or that the black hole has a larger 
entropy capacity than usually believed, or that quantum effects allow information to escape 
from the black hole despite the classical horizon, or that the region behind the horizon does not exist,
or that information is hidden in ``parallel worlds''.
Additional insight is needed to distinguish the proposals which are 
promising from those which are not. 

The standard mechanism of Hawking radiation based on semiclassical approximation \cite{hawk1} is 
certainly not completely correct, so one might think that a key to the resolution of the paradox 
is to better understand the mechanism of Hawking radiation itself. 
In this paper, however, we find a new version of b.h. information paradox
in which Hawking radiation does not play any important role. This suggests 
that the mechanism of Hawking radiation as such is not essential for the paradox resolution.

Our basic idea is a simple thought experiment consisting of only two steps. 
In the first step an entangled pair of low-energy photons is prepared, 
and in the second step one member of the pair is dropped into the black hole. 
In this way one may significantly increase the b.h. entropy 
without a significant increase of its mass. This is a paradox, 
because it seems to violate all reasonable entropy bounds.
  
This new version of the information paradox 
is logically independent from the standard one
and may be interesting on its own. Nevertheless, our main motivation for studying it
is to shed some new light on the standard version of the paradox. 
It turns out that most of the proposed solutions to the standard b.h. information paradox
do not help to solve the new version, suggesting that the new version of the information paradox is even a harder 
problem than the standard one. However, we turn this hardness into a virtue, by proposing that both
versions of the paradox should be solved in the same manner. In this way, only a few of the proposed resolutions
to the standard b.h. information paradox survive as promising approaches to solve both problems at once.

In the rest of the paper we elaborate those ideas in more detail.

\section{Entropy with (almost) no energy}

Thermal entropy $S_{\rm th}$ is necessarily associated with energy through the thermodynamic relation $TdS_{\rm th}=dQ$, 
where $T$ is temperature and $Q$ is heat. However, thermal entropy is not the only kind of entropy. 
Namely, thermal entropy is entropy of systems in (or close to) thermal equilibrium, while many physical systems 
with entropy are far from thermal equilibrium.  
With a non-thermal entropy at hand, one can, in principle, increase entropy of a system without increasing its energy.

The most interesting example is the entanglement entropy, 
which can characterize thermal and non-thermal systems.
Let $|e,1\rangle$ and $|e,2\rangle$ be two orthogonal quantum states 
with the same energy $e$. Taking them to be the states of some massless field such as the electromagnetic field,
the energy $e$ can, in principle, be arbitrarily small. In the simplest case $|e,1\rangle$ and $|e,2\rangle$ are
one-particle states, but in principle they can also be many-particle states. From such states one can prepare 
an entangled state
\begin{equation}\label{entang}
\frac{1}{\sqrt{2}} [ |e,1\rangle_{\rm L}|e,2\rangle_{\rm R} + |e,2\rangle_{\rm L}|e,1\rangle_{\rm R} ]
\end{equation}
where ${\rm L}$ and ${\rm R}$ denote two spatially separated subsystems, one positioned at the ``left'' side 
of the laboratory and the other at the ``right''. The subsystem on the left is a mixed state
described by the density matrix
\begin{equation}\label{rho}
 \rho_{\rm L}=\frac{1}{2} (|e,1\rangle\langle e,1|)_{\rm L}+\frac{1}{2} (|e,2\rangle\langle e,2|)_{\rm L} .
\end{equation}
This subsystem has energy $e$ and the entanglement entropy $S=\ln 2$.
In principle, as we said, $e$ can be arbitrarily small. 
(The energy of $n$ photons of frequency $\omega$ is
$e=n\hbar\omega$. Even though $n$ is integer, this energy can be 
arbitrarily small because $\omega$ can be arbitrarily small.)
Hence it makes sense to consider the limit $e\rightarrow 0$,  
which is a subsystem that carries entropy without carrying energy.
 
Note that the wave function of a massless particle with energy $e$ is a wave packet of the spatial width 
$W\:\, {\raise-.5ex\hbox{$\buildrel{{\textstyle \!>}}\over{\!\sim}$}} \, e^{-1}$. 
(Unless stated otherwise, we work in units $\hbar=c=k_{B}=G_{N}=1$). This means that, in Minkowski spacetime,
such a low-energy entropy cannot be packed into a small box. We shall see that it changes drastically when a box 
in Minkowski spacetime is replaced by a black hole. 

\section{Dropping entropy into the black hole}

\subsection{Theoretical aspects}

Now suppose that the subsystem on the left is dropped into the black hole of initial mass $M$.  
Initially, the subsystem on the left is far from the black hole and has a large size 
$W\:\, {\raise-.5ex\hbox{$\buildrel{{\textstyle \!>}}\over{\!\sim}$}} \, e^{-1}$.
If $e^{-1}$ is larger than the b.h. radius $R=2M$, one might naively think that 
the wave packet on the left cannot be inserted into the black hole. However, as the wave packet approaches 
the b.h. horizon, it suffers the exponential blueshift. From the point of view of the external observer
far from the black hole, this means that the wave packet shrinks by an exponential factor and becomes
much smaller than the b.h. radius $R$. Hence the black hole can absorb 
the subsystem on the left, so the b.h. entropy and mass are increased by 
\begin{equation}\label{deltaSe}
\delta S=\ln 2, \;\;\;\;  \delta M=e.
\end{equation}

Now let us see how it can be used to violate the Bekenstein bound. To simplify the analysis, we shall assume that
the black hole does not have angular momentum and electric charge, and that the states $|e,1\rangle$ and $|e,2\rangle$
have a zero total angular momentum, ${\bf J}={\bf L}+{\bf S}=0$. Then, according to the Bekenstein bound \cite{bekenstein81},
the b.h. entropy $S_{bh}$ cannot be larger than the Bekenstein-Hawking entropy 
\begin{equation}\label{SBH}
 S_{BH}=\frac{A}{4}=4\pi M^2, 
\end{equation}
where $A=4\pi R^2$ is the b.h. area. Taking the differential of (\ref{SBH}) one gets 
$dS_{BH}=8\pi MdM$, which implies that the Bekenstein bound can be violated if
\begin{equation}\label{dS}
\delta S> 8\pi M\delta M .  
\end{equation}
Using (\ref{deltaSe}), this implies that the Bekenstein bound can be violated if
\begin{equation}\label{e}
 e<\frac{\ln 2}{8\pi} \frac{1}{M}=\frac{\ln 2}{8\pi} \frac{m_{\rm Pl}}{M} m_{\rm Pl} ,
\end{equation}
where in the last equality we restored the units in which the Planck mass $m_{\rm Pl}=G_{\rm N}^{-1/2}$
is not unit. No known physical law forbids preparation of quantum states with energy satisfying (\ref{e}).
Indeed, for $M$ which is small by astronomical standards (say, $M\sim 10^9 m_{\rm Pl}$), 
a state of photon(s) with energy $e$ satisfying (\ref{e}) can easily be prepared and manipulated with current 
quantum-optics technologies. 
  
Conceptually, such a violation of the Bekenstein bound is somewhat similar to the Bekenstein-bound violation
by the monster states studied in \cite{monster}. However, the formation of monster states studied in \cite{monster}
requires rather unrealistic initial conditions. Our mechanism for Bekenstein-bound violation
requires only an ordinary initial black hole and some quite realistic manipulations of photons.
For other mechanisms of Bekenstein-bound violation see also \cite{page-bek} and references therein.
 
For the sake of nit-picking, it may also be relevant to distinguish two different interpretations of the Bekenstein bound.
In one interpretation, a black hole should {\em always} have the entropy equal to (\ref{SBH}).
So if, indeed, the initial b.h. entropy is equal to (\ref{SBH}), the Bekenstein bound will be violated
by preparing only one entangled pair (\ref{entang}) followed by the subsequent dropping. 
In another interpretation, the initial b.h. entropy may be smaller than (\ref{SBH}). In this interpretation
one entangled pair (\ref{entang}) may not be sufficient to violate the Bekenstein bound,
but it does not make much difference because one can prepare a large number $N$ of independent
entangled pairs of the form (\ref{entang}), and then one can drop into the black hole the left member of each pair.
In this way the entropy and mass increase by $\delta S=N\ln 2$ and $\delta M=Ne$, so the 
Bekenstein bound can always be violated by a sufficiently large $N$. In both interpretations,
the only crucial condition for the Bekenstein-bound violation is Eq.~(\ref{e}). 

If violation of the Bekenstein bound is not surprising enough, we point out that in a similar way
any other reasonable bound can be violated. For instance, if the maximal b.h. entropy scales with
``naive'' volume $4\pi R^3/3$, or with the much larger internal volume \cite{rovelli_vol}, 
one can always find a sufficiently small energy $e$ that violates that bound. 
In general, any reasonable entropy bound of the form
\begin{equation}\label{Sbh}
S_{\rm bh}\le f(M) 
\end{equation}  
with $f'(M)\equiv df(M)/dM>0$ can be violated by choosing 
\begin{equation}\label{e3}
 e<\frac{\ln 2}{f'(M)}.
\end{equation}
Eq.~(\ref{e}) is nothing but a special case of the general condition (\ref{e3}).


\subsection{Practical aspects}

The blueshift discussed above shrinks a wave packet in the longitudinal direction, but not in the transverse direction.
A low energy wave packet which in the transverse direction is initially much wider than the b.h. radius $R$ 
will typically remain so when the wave packet approaches the black hole. Hence, in practice, it is not so easy
to achieve the wave packet absorption by the black hole. Such a problem does not appear for a theoretical black hole
in $1+1$ dimensions, but it appears for realistic black holes in $3+1$ dimensions. Here we discuss various possibilities
to resolve this practical problem.

\begin{enumerate}

\item {\it Fine tuning of initial conditions.} 
Hawking radiation can be described in terms of wave packets \cite{hawk1} that are well localized near the horizon 
initially, but widely spread around the black hole in the final sate. The Schwarzschild black hole is time-independent,
so the wave equation in the Schwarzschild background is time-inversion invariant. Hence there 
exist time-inverted solutions of the wave equation that describe wave packets which are widely spread around the black hole
initially but localized near the horizon when the packet approaches the horizon. No known principle of physics 
forbids preparation of such wave packets, which makes the wave packet absorption possible in principle.
In practice, however, a preparation of such a wave packet would require a fine tuning of initial conditions,
suggesting that this approach might be too difficult in practice.

\item {\it Engineering tricks.} 
Instead of dealing with fine tuning of initial conditions, one can devise various engineering tricks that can 
help the wave packet to enter the black hole. 
One possibility is to build a wave guide shaped such that it is wide far from the horizon but
narrow near the horizon. 
Another possibility is to build the quantum-optics laboratory near the horizon,
so that the wave packet is localized near the horizon initially. With this second possibility the photons may have
a much larger initial energy as measured by observers in the laboratory, 
because their contribution to the b.h. mass is determined by the 
small red-shifted energy measured by observers far from the black hole. 
Of course, keeping the wave guide or the laboratory at a stable position near the horizon may lead to additional
practical problems, but in principle such problems are not insurmountable.

\item {\it Many trials.}
An atom can absorb a photon the wave length of which is much larger than the size of the atom \cite{ballentine}.
The catch, of course, is that such absorption is not a classical deterministic process. The absorption 
is a quantum ``jump'', the probability of which is very small. A black hole can absorb a soft photon in the same sense,
the probability for which is very small because the cross section of photon scattering on a black hole 
is of the same order of magnitude as the b.h. area \cite{das-gibbons-mathur}. 
The small probability $p$ of absorption 
of a single copy of (\ref{rho}) can be overcame by a large number $N_{\rm tr}$ of trials, with each trial performed 
with another copy of (\ref{rho}). In this way the number $N$ of successful trials for which the photons are actually absorbed is 
\begin{equation}
 N\sim pN_{\rm tr} ,
\end{equation}
which can be made arbitrarily large by taking a sufficiently large $N_{\rm tr}$. For practical purposes, the method
with many trials is probably the best. 

\end{enumerate}

\section{Resolution of the paradox - useless approaches}

The conclusion that any reasonable entropy bound can be violated by dropping sufficiently soft 
massless particles into the black hole is a paradox, so it needs to be resolved. 
Can this paradox be resolved in the same way as the standard b.h. information paradox with Hawking radiation?
In the next section we shall discuss which approaches to the standard information paradox may also be useful 
for the resolution of our version of the paradox. 
In this section we shall first eliminate those approaches that do {\em not} seem useful for our paradox.

\begin{enumerate}

\item {\it No Hawking radiation.} 
One logical possibility is that Hawking radiation does not exist \cite{helfer}.
While it obviously avoids the standard b.h. information paradox, it does not help because our paradox does 
not depend on the existence of Hawking radiation. 

\item {\it New physics for small black holes.} 
Proposals of that sort include Planck-sized remnants 
\cite{rev_remn}, creation of a baby-universe \cite{gid} and sudden escape of information, perhaps via tunneling into
a white hole \cite{rovelli_tunnel}. Presumably all such events happen when the black hole becomes sufficiently small,
which is useless for our purpose because our version of the information paradox exists also for large black holes.

\item {\it Mild modifications of horizon physics.} 
One possibility is that quantum fluctuations at the horizon allow a slow leak of information \cite{gid}.
Another possibility is that prehawking radiation prevents creation of an apparent horizon \cite{kawai,terno}.
Since such scenarios involve rather slow processes (slow leaking or slow prehawking radiation in the examples above),
they are essentially useless because their effect may easily be overpowered by 
dropping many copies of (\ref{deltaSe}) at almost the same time.

\item {\it Radical modifications of horizon physics.}
It has been proposed that quantum gravity effects make the b.h. horizon totally impenetrable, due to
a fuzzball \cite{mathur}, an energetic curtain \cite{braunstein} or a firewall \cite{AMPS}
at $R=2M$. How such an impenetrable barrier could act in an attempt to drop (\ref{deltaSe}) into the hole? 
If the dropped particles would accumulate in a small region in front of the wall,
that would be useless because (\ref{e3}) would again violate any reasonable
entropy bound in that small region.
Alternatively, if the dropping would result in a fast recoil of the dropped particles so that
the particles cannot accumulate near $R=2M$, that would resolve our version of the b.h. information paradox.
However, such a recoil would be observed in astrophysical
black holes such as the one in the center of our galaxy, which is not what we observe.
One might argue that we don't observe it yet because the firewall forms only after a very long time 
(Page time \cite{apologia}), but then we are back to the problem that fast dropping of entropy can
violate any reasonable entropy bound, much before the firewall forms. 
 
\item {\it Complementarity.}
Even though quantum cloning contradicts unitarity, according to the b.h. complementarity principle \cite{complementarity}
it is acceptable as long as no single observer can see both copies. This means that 
one copy of (\ref{rho}) can be destroyed in the black hole, while the other copy can remain outside of the black hole.
The outside copies must either be accumulated near $R=2M$ or recoiled, leading to the same problems 
as with the firewall above.  

\item {\it Decoherence and many worlds.}
Radiation of a single Hawking particle is a random quantum event. The particle energy can take any value
from a large range of possible values. Unitarity, combined with decoherence
induced by the macroscopic environment, 
implies that the total wave function of the universe contains all the branches corresponding to the all 
possible energies of Hawking particles. While it may help to resolve the standard b.h. information paradox
\cite{kiefer,zeh,nik_bhstring,nomura,hsu,hollowood,carroll_bhbranch},
here it is useless because the states $|e,1\rangle$ and $|e,2\rangle$ in (\ref{rho}) have the same energy
and can be chosen to be indistinguishable at the macroscopic level. This means that the macroscopic environment cannot 
distinguish $|e,1\rangle$ from $|e,2\rangle$ and therefore cannot create different branches.

\item {\it Soft hair.}
It has been argued that black hole has infinitely many soft supertranslation hair, so that
Hawking radiation can be entangled with that hair \cite{softhair}.
This can help to resolve the standard b.h. information paradox, but here it is useless because
it does not influence our mechanism for violation of entropy bounds before the evaporation.

\end{enumerate}

\section{Resolution of the paradox - potentially useful approaches}

Now let us discuss those approaches to the standard  b.h. information paradox that can also resolve
our version of the paradox.

\begin{enumerate}

\item {\it Information destroyed in the singularity.}
If, as originally proposed by Hawking \cite{hawk2}, any excess of information induced by Hawking radiation is 
destroyed in the b.h. singularity, then so is the any excess of information dropped into the black hole by our
mechanism. In this sense, information destruction in the singularity is probably the simplest resolution 
of our version of the paradox. 
It has been argued that such a non-unitary evolution violates energy-momentum conservation or locality \cite{BSP},
but a more careful analysis reveals that it is not the case for systems with a large number of degrees of freedom \cite{nik_nounit}.
Moreover, by treating time as a local quantum observable, such information destruction can be reinterpreted
as a unitary process in disguise \cite{hartle1,hartle2,nik1,nik2,nik3}.

\item {\it ER=EPR and islands.}
According to the ER=EPR conjecture \cite{er=epr}, the left and right subsystems in (\ref{entang})
are connected by a wormhole. Therefore, instead of being destroyed in the singularity at $r=0$,
the left subsystem can escape from the black hole through the wormhole. 
Such an escape that bypasses the horizon resolves our version of the b.h. information paradox.
A more precise version of this idea involves a black hole island \cite{island}, a region in black hole
that due to a wormhole should be thought of as a part of the b.h exterior, rather than interior.

\item {\it Gravitational crystal.}
By analogy with condensed-matter physics, it has been proposed that general relativity is merely a macroscopic description 
of a fluid phase of some unknown fundamental degrees of freedom \cite{nik_cryst}. Those fundamental degrees can also exist
in the crystal phase that does not obey the laws of general relativity.
Instead of being destroyed in the singularity at $r=0$, any excess of information in the black hole
gets absorbed by a crystal core formed around the center at $r=0$. 
The entropy of the core scales with its volume $V_{core}$, so, instead of (\ref{Sbh}),
the relevant entropy bound is $S_{core}\le \alpha V_{core}$ with $\alpha\sim 1$.
Consequently, the core continuously grows as new information arrives \cite{nik_cryst}.
In this way, since general relativity is not valid in the crystal phase, the core can penetrate 
the horizon from the inside and become even larger than $R=2M$.
 

\end{enumerate}

\section{Discussion and conclusion}

By dropping entanglement entropy of low-energy massless particles into the black hole
one can violate any reasonable entropy bound of the form (\ref{Sbh}), 
which constitutes a new version of the b.h. information paradox. Unlike the standard version of the paradox \cite{hawk2}, 
the new version does not depend on the existence of Hawking radiation.
We have argued that most of the known proposals for the resolution of the standard version of the
paradox are not useful for a resolution of this new version. 
The only proposals for the standard version of the paradox that we found useful for the new version 
are the information destruction \cite{hawk2}, ER=EPR/islands \cite{er=epr,island},
and the gravitational crystal \cite{nik_cryst}.
However, it does not necessarily imply that all other proposals should be rejected. Perhaps some of the eliminated
resolutions can still be useful in some refined form,
or perhaps the new version of the paradox should be resolved by ideas
that do not depend on the resolution of the standard version.
In any case, we believe that our new version of the paradox offers a new insight that 
may stimulate further fundamental research on the b.h. information paradox.

\section*{Acknowledgements}
The author is grateful to L. Maccone for useful discussions.
This work was supported by 
the Ministry of Science of the Republic of Croatia. 

\end{document}